# A CONSTANT SELF-CONSISTENT SCATTERING LIFETIME IN SUPERCONDUCTING STRONTIUM RUTHENATE

Pedro L. Contreras E.

Address: Departamento de Fisica, University of the Andes, 5101, Mérida, Venezuela
Corresponding author email: pcontreras@ula.ve

## ABSTRACT

In this numerical work, we find a self-consistent constant scattering superconducting lifetime for two different values of the disorder parameters, the inverse atomic strength, and the stoichiometric impurity in the triplet paired unconventional superconductor strontium ruthenate. This finding is relevant for experimentalists given that the expressions for the ultrasound attenuation and the electronic thermal conductivity depend on the superconducting scattering lifetime, and a constant lifetime fits well nonequilibrium experimental data. Henceforth, this work helps experimentalists in their interpretation of the acquired data. Additionally, we encountered tiny imaginary parts of the self-energy that resembles the Miyake-Narikiyo tiny gap outside the unitary elastic scattering limit, and below the threshold zero gap value of 1.0 meV.

**Keywords**: Constant superconducting lifetime, Strontium ruthenate, Unconventional superconductivity, Ultrasound attenuation, Electronic thermal conductivity. Self-consistent method.

## INTRODUCTION

This work aims to understand how the scattering lifetime of the triplet paired unconventional superconductor strontium ruthenate [1-6] is constant if calculated self-consistently in the presence of disorder. The existence of some kind of disorder in strontium ruthenate was proposed years ago [7]. A numerical answer appears by adding nonmagnetic disorder and scanning properly, the values of the zero superconducting gap $\Delta_0$, the inverse atomic strength, $c$, and the stoichiometric disorder concentration $\Gamma^+$ [8] using a self-consistent procedure. An elastic scattering lifetime is an essential part for the analysis of nonequilibrium superconducting ultrasound attenuation, and electronic thermal conductivity [9-14]. Therefore, we consider the use of the self-consistent elastic cross-section, a robust and powerful tool because it shows several hidden physical features as the presence of tiny gaps or point nodes [15] due to the anisotropic parameterization with the same order parameter (OP) if it breaks the time-reversal symmetry as it happens in $Sr_2RuO_4$.

The original proposal of a constant superconducting lifetime in the nonequilibrium properties comes from [16]. Furthermore, experimental, theoretical and numerical studies of the kinetic coefficient using the Green function formalism is a well-developed approach [9-14], but there was not understanding of why superconducting $\tau_s$ has to be constant in order to fit experimental data in some alloys as it is outline below with the respective references. Moreover, we understand from this work that the difficulty dwells in the self-consistent calculation of the imaginary cross-section term, problem which not only belongs to the quantum mechanical nonrelativistic scattering theory, but also to the nonequilibrium statistical mechanics.

We mention, three equations where the calculation of the scattering lifetime comes out. First, for the distribution function $f_B(t)$ in the Boltzmann equation [17]. Second, for the quasiclassical probability density $W(t)$ calculated using nonequilibrium statistical mechanics [18]. Third, for the characteristic time equation in a chemical reaction involving a reactant [A](t) [19]. In chemical kinetics, the self-consistent task is called the stiffness problem [19], and it happens when the dependent variable changes very fast. Table 1 summarizes the three different nonequilibrium equations with a self-consistent lifetime problem, and for the first two, the solution is a challenge prompt beyond the $\tau$−approximation for electrons [20].

Table 1: Kinetic equations with a self-consistent lifetime.

| Boltzmann equation for the distribution function $f(t)$ [17] | Probability density equation $\mathcal{W}(t)$ [18] | Chemical reaction equation, and its characteristic lifetime [19] |
|---|---|---|
| $\left(\partial f_B(t) / \partial t\right)_{coll} + 1/\tau\, f = 0$ | $\left(\partial \mathcal{W}(t) / \partial t\right)_{qsd} + 1/\tau\, \mathcal{W} = 0$ | $d\,[A](t) / d\,t + 1/\tau\,[A] = 0$ |

In this section, we briefly introduce one nonequilibrium property that fits well data if a constant scattering lifetime is used, i.e., the superconducting ultrasound attenuation. An experimental ultrasound data in strontium ruthenate [21] was fitted successfully [22] using the directional superconducting ultrasound attenuation equation that comes from the Green function formalism

$$\alpha_j(T) \Big/ \alpha_j(T_c) = 1 \Big/ (2\,T) \times 1 \Big/ \langle f_j^2 \rangle_{FS} \times \int_0^\infty d\epsilon \, A(\epsilon) \ (1),$$

In equation (1) the energy term is  $A(\epsilon) = \left[ cosh^2 \left( \frac{\epsilon}{2\,T} \right) \right]^{-1} \frac{\tau_s(\epsilon)}{\tau_n} \frac{\langle f_j^2 \times \Re \sqrt{\epsilon^2 - \Delta_k^2} \rangle_{FS}}{\epsilon}$ .

Function $A(\epsilon)$ shows explicitly the ratio of interest in this work $\tau_s(\epsilon) \big/ \tau_n$ . Equation (1) contains other two terms averaged over the Fermi surface: First, the square of the electron-phonon matrix elements $f_j$, for the three elastic modes "$j = P, T_1,$ and $T_2$". Second, the real part of the root $\sqrt{\epsilon^2 - \Delta_k^2}$ times the square of  $f_j$, i.e.,$\langle f_j^2 \times \Re \sqrt{\epsilon^2 - \Delta_k^2} \rangle_{FS}$. In addition, $A(\epsilon)$ has the trigonometric factor that is the negative derivate of the Fermi-Dirac distribution $[cosh^2 \left( \frac{\epsilon}{2\,T} \right)]^{-1} = -\frac{\partial f_F}{\partial \epsilon}$, the temperature $T$, and the energy $\epsilon$ (self-consistency is carried by this variable). A detailed derivation of equation (1) in the general case is given in the appendix of [23]. Other works also reported the use of a constant scattering lifetime [24], meanwhile others researchers used a Fourier OP expansion to fit the experimental data in strontium ruthenate [25].

The ultrasound attenuation term $(\alpha)$ in equation (1) comes out from the imaginary part of the polarization operator [26,27], this expression has been used to study the anisotropic gap directional structure in conventional [28,29] and unconventional superconductors. It should be pointed out that is instructive to compare with the UPt$_3$ heavy fermion compound [30-32] due to the evidence found from ultrasound studies of several superconducting phases, and the complicated structure of its Fermi surface. On the other hand, other properties worth to mention from the interaction of electrons and sound waves in strontium ruthenate are the splitting of the critical temperature in an uniaxial external stress [33,34], the unusual power law behavior of the electron-phonon interaction in the $\alpha, \beta$ and $\gamma$ Fermi surfaces sheets below the transition temperature $T_c$ [21,22], the unusual behavior of the normal state viscosity above $T_c$ [21,23], and the temperature power law activity due to different phonon modes [35].

The first unconventional compound to show a jump at $T_c$ in one elastic velocity mode was the heavy fermion alloy UPt$_3$. This physical property was interpreted as a signature of a broken symmetry [36,37]. In the case of Sr$_2$RuO$_4$ the first experimental discover that the time reversal symmetry was broken is due to [3]. Theoretically, the velocity jump using the Green function formalism is given by the real part of the polarization operator [27]. However, several uniaxial stress $\epsilon_{ij}$ experiments and calculations using the Gibbs thermodynamic potential $G(T)$ with an external elastic field  have been performed for an elastic broken symmetry field in strontium ruthenate [38-42].

The other nonequilibrium property, i.e., the electronic thermal conductivity $(\kappa)$in the superconducting strontium ruthenate was experimentally measured [43] and numerically calculated in [44] giving an excellent agreement with the fit performed. In this case the use of a constant superconducting lifetime ratio was a key point for the successful fitting. Therefore, the reason for a constant $\tau_s$ in this compound needs further research and it is addressed in this work. It is worth mentioning, that for $\kappa$, two signatures are relevant, the universal behavior at very low temperatures, and the reduction of the transition temperature with nonmagnetic disorder [45,46].

In the following two sections, we report a numerical analysis supporting our findings, i.e., flats imaginary parts of the elastic scattering cross-section, concluding that a constant self-consistent superconducting lifetime is possible in strontium ruthenate for two different concentration of stoichiometric disorder, i.e., when the disorder is dilute and near the unitary limit, and for enriched disorder in the intermediate scattering region.

## NUMERICAL PROCEDURE

The self-consistent Green-functions treatment of the inverse scattering lifetime $\tau_s$ using rationalized Planck units ($\hbar = k_B = $ c $= 1$), where c is the speed of light is performed using the expresion [14]

$$1 \Big/ \tau_s = 2\,\Im\,[\widetilde{\omega}(\omega + i\,0^+)] \ (2),$$

Equation (2) connects $\tau_s$ and the imaginary part of the elastic cross-section (we abbreviate it as $\Im[\widetilde{\omega}]$). In strontium ruthenate the normal state $\tau_n$ is constant, since the compound is a robust Fermi liquid with three sheets [47]. A normal metal follows a simple lifetime equation $1/\tau_n = 2\,C$, with a constant value $C$. Henceforth, the ratio $\tau_s/\tau_n$ becomes $\tau_s(\widetilde{\omega})/\tau_n = C/\Im[\widetilde{\omega}]$. Although, if the scattering lifetime also becomes constant in the superconducting state, we have that $1/\tau_s = 2\,\Im[\widetilde{\omega}] = C'$, and the normal to superconducting lifetime ratio (abbreviated as NS) is a single constant value $\tau_s/\tau_n = C/C' = \mathbb{C}$.

Nevertheless, it is physically significant to find as well, where the constituents quasiparticles of strontium rutenate scatters mostly, i.e. in the hydrodynamic limit, in the intermediate region, in the unitary limit, or in between any of these elastic scattering phases, as we further explain in the next section.

To obtain the numerical behavior linking the lifetime with nonmagnetic disorder, the expression of the imaginary part of the cross-section (also, half of the inverse scattering lifetime $(2\,\tau)^{-1}$, and half of the collisional frequency $\nu/2$), the following relation holds $1/2\tau = \nu/2 = \Im[\widetilde{\omega}]$. In this way, $C$ is the constant inverse value of the lifetime in the normal state, $C'$ is the constant inverse value of the lifetime in the superconducting state, $\mathbb{C}$ is the NS ratio, c is the speed of light in rationalized Planck units, and $c$ is the dimensionless inverse strength atomic parameter, $c \sim U_0^{-1}$.

We use a group theoretical expression for a triplet gap that breaks time-reversal symmetry that is consistent with the experiments mentioned for strontium ruthenate, the Miyake-Narikiyo triplet order parameter [48]. it is worth mentioning other group theoretical models used to explain the nodal structure of Sr2RuO4, those models have an OP with line nodes, points nodes, or combinations of them, and also fit successfully some experimental data as the superconducting specific heat [49-52].

The behavior of the pair of coordinates given by $(\Re[\widetilde{\omega}], \Im[\widetilde{\omega}])$ for strontium ruthenate is analyzed in the reduced phase space (RPS) by sketching a phenomenological phase diagram [53]. The RPS has two coordinates, the axis 0X is the real part of the self-consistent frequency $\Re[\widetilde{\omega}]$, and the axis 0Y its imaginary term $\Im[\widetilde{\omega}]$. Besides, our numerical procedure controls five input physical parameters in the anisotropic case: the Fermi level $\epsilon_F$, the first tight binding hoping parameter $t$, the zero gap $\Delta_0$, the inverse scattering strength $c$, and the stoichiometric strontium disorder $\Gamma^+$.

As was pointed out in the introduction, strontium ruthenate is a triplet superconductor with $T_c \approx 1.5$ K and depending on disorder. It belongs to the complex 2D irrep OP E$_{2u}$ [33,42]. The structure of the gap is $\boldsymbol{\Delta}(k_x, k_y) = \Delta_0\,\boldsymbol{d}(k_x, k_y)$, with the vector $\boldsymbol{d}(k_x, k_y) = \big(\sin(k_x a) + i\,\sin(k_y a)\big)\hat{\boldsymbol{z}}$, and $\Delta_0$ is the zero gap parameter that in this work take values from zero meV in the normal state, to the threshold value equal to 1.0 meV.

The numerical solution contains the dependent self-consistent variable $\widetilde{\omega}$ that changes very fast [19]. In addition $\widetilde{\omega}$ is found on both sides of the equation (3), thus $\widetilde{\omega}$ equals to [54,55]

$$\widetilde{\omega}(\omega + i\,0^+) = \omega + i\,\pi\,\Gamma^+\,\frac{g(\widetilde{\omega})}{c^2 + g^2(\widetilde{\omega})}\,(3).$$

Numerical disorder is controlled in the imaginary part of equation (3) with the help of two parameters, $c$ and $\Gamma^+$ [54,55]. The dimensionless inverse strength parameter $c = (\pi\,N_F\,U_0)^{-1}$, where $U_0$ is an impurity atomic potential and $N_F$ is the density of states at the Fermi level. The other parameter is the strontium stoichiometric concentration with $\Gamma^+ = n_{dis}/(\pi^2\,N_F)$. The function $g(\widetilde{\omega})$ is given by the expression $g(\widetilde{\omega}) = \langle \frac{\widetilde{\omega}}{\sqrt{\widetilde{\omega}^2 - |\Delta|^2(k_x, k_y)}} \rangle_{FS}$.

The use of the isotropic Fermi surface approach to equation (3) with the disorder parameters $c$ and $\Gamma^+$ is summarized in [56], but we can use a tight-binding model as done for the HTSC strontium-doped lanthanum cuprate [57] and strontium ruthenate [8]. In the case of the strontium ruthenate it is found with the same OP a quasipoint nodes or a point nodes structures by only changing the value of the Fermi level [15]. In this work, the Fermi level is $\epsilon_F = 0.4$ meV and the first neighbors hoping parameter is $t = 0.4$ meV. The normal state energy is $\xi(k_x, k_y) = -+2\,t\,[\cos(k_x a) + \cos(k_y a)]$. The Fermi level value $\epsilon_F = 0.4$ meV makes the normal state energy anisotropic, and leaves a quasipoint tiny gap around the $(0, \pm\pi)$ and $(\pm\pi, 0)$ cell points [48]. $\boldsymbol{d}(k_x, k_y)$ has two components which belong to the irreducible representation $E_{2u}$ of the tetragonal point group $D_{4h}$. It corresponds to a triplet odd paired state $\boldsymbol{d}^y(-k_x, -k_y) = -\boldsymbol{d}^y(k_x, k_y)$ with a complex irreducible representation composed by the functions $\sin(k_x a)$ and $\sin(k_y a)$ and the Ginzburg-Landau coefficients $(1, i)$ [42].

If $c = 0.0$, equation (5) is in the unitary limit $\Im[\tilde{\omega}] = i\,\pi\,\Gamma^+ \frac{1}{g(\tilde{\omega})}$ where $\Gamma^+$ is kept as a unique parameter. However, in this particular self-consistent simulation, we use two expressions: The subsection one of the next section is computed with a dilute disorder imaginary expression $\Im[\tilde{\omega}] = 0.05\,\pi\,meV\,\frac{g(\tilde{\omega})}{(0.2)^2 + g^2(\tilde{\omega})}$ (between the unitary and intermediate regions). Subsection two of the next section is calculated with an intermediate scattering region, and an enriched disorder value, i.e., $\Im[\tilde{\omega}] = 0.2\,\pi\,meV\,\frac{g(\tilde{\omega})}{(0.4)^2 + g^2(\tilde{\omega})}$.

In what follows, simulations for the zero gap are performed in this case with the parameters $c = 0.2$, $c = 0.4$, a dilute stoichiometric disorder $\Gamma^+ = 0.05$ meV, and the enriched stoichiometric disorder $\Gamma^+ = 0.20$ meV. Other values of $\Gamma^+$ for $c = 0.2$ are also compared, $\Gamma^+ = 0.10$ meV and $\Gamma^+ = 0.15$ meV, both of them considered in the range of an optimal disorder. Additionally, we use eleven values for the zero superconducting gap $\Delta_0$ in each case, starting from the normal state ($\Delta_0 = 0.0$ meV), adding a 0.1 meV to each zero gap value. The threshold limit is the experimental value for strontium ruthenate ($\Delta_0 = 1.0$ meV), as this value shows a single tiny gap in the unitary limit.

## NUMERICAL RESULTS

### 1. Dilute disordered constant superconducting lifetime

Figure 1 shows the behavior of dressed fermionic carriers for which a constant scattering lifetimes is observed in two regions of the reduced phase space. The first interval is the normal state phase with real frequencies between ($\pm 3.1$, $\pm 4.0$) meV and a constant imaginary frequency $\Im[\tilde{\omega}] = 0.149$ meV. The second interval is localized inside the superconducting phase and shows a superconducting scattering lifetime constant for real frequencies in the interval $(0.00, 0.26)$ meV (see Table 2 for a numerical sample of the numbers in the superconducting case). The superconducting zero gap parameters in Fig. 1 correspond to the values $\Delta_0 = 0.0$ (normal state), 0.1, 0.2, 0.3, 0.4, 0.5, 0.6, 0.7, 0.8, 0.9, and 1.0 (threshold value) meV

Since $\Im[\tilde{\omega}]$ is constant, it corresponds to a constant scattering lifetime in the superconducting state of strontium ruthenate, confirming the general prediction for unconventional superconductors [16], and the fittings of experimental data for $\alpha$ [21,22] and $\kappa$ [43,44] We show in Table 2, a sample of the real, and the imaginary RPS coordinates with the flat shape in Fig. 1. This happens for $c = 0.2$, that means outside the unitary and the intermediate limits, with a dilute stoichiometric disorder value given by $\Gamma^+ = 0.05$ meV. Three microscopic zero superconducting gap values are shown in Table 2: $\Delta_0 = 0.8, 0.9$, and 1.0 (meV).

From Fig. 1 and Table 2, we find that the superconducting $\Im[\tilde{\omega}]$ has a value of 0.362 meV, bigger than the value $\Im[\tilde{\omega}] = 0.149$ meV in the normal state. That means that the constant scattering lifetime is smaller in the superconducting dilute disorder phase. The superconducting collisional frequency is bigger than in its normal state, probably due to the existence of scattering processes coming from other dressed fermions, and the phonon quasiparticles (sound waves) with the strontium disordered atoms. We think that for the unconventional strontium ruthenate at low frequencies, the momentum of the dressed quasiparticles is transferred to strontium stoichiometric disordered atoms in the crystal lattice, contrary to what happens in the high $T_c$ lightly doped lanthanum cuprate La2-xSrxCuO4, where the $\boldsymbol{k}$-transfer happens at higher or much higher frequencies in the unitary limit and it becomes harder to find the frequency point of the phase transition in the RPS [58].

A simple physical picture tell us that Sr atoms migrate through the lattice, and stick together in a coalescing metallic state with a constant lifetime, but this time it happens for a dilute disorder of stoichiometric Sr atoms ($\Gamma^+ = 0.05$ meV) and low frequencies (if a species of an atom is stoichiometric, an infinite unitary $U_0$ is not needed to form a coalescent metallic phase).

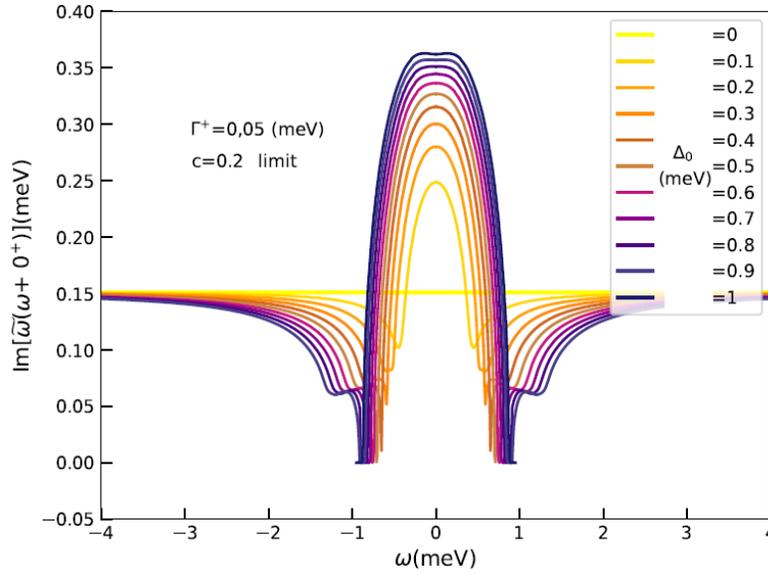

Fig. 1: Evolution of the zero superconducting gap for c = 0.2 and dilute disorder in strontium ruthenate. This plot is flat at real frequencies interval (0.00, 0.26) meV. Also, notice several tiny gaps for $\Delta_0 = 0.8, 0.9,$ and 1.0 meV.

Table 2: Sample of real & imaginary RPS frequencies and different zero gap (all values are in meV units) for $c = 0.2$ with dilute disorder:

| $\omega = \Re(\widetilde{\omega})$ | 1.00e-003 | 2.10e-002 | 4.10e-002 | 7.10e-002 | 1.01e-001 | 1.31e-001 | 1.51e-001 | 1.81e-001 | 2.61e-001 |
|---|---|---|---|---|---|---|---|---|---|
| $\Im(\widetilde{\omega})$ $\Delta_0 = 1.0$ | 3.62e-001 | 3.62e-001 | 3.62e-001 | 3.62e-001 | 3.62e-001 | 3.62e-001 | 3.62e-001 | 3.62e-001 | 3.60e-001 |
| $\Im(\widetilde{\omega})$ $\Delta_0 = 0.9$ | 3.57e-001 | 3.57e-001 | 3.57e-001 | 3.57e-001 | 3.57e-001 | 3.57e-001 | 3.56e-001 | 3.56e-001 | 3.55e-001 |
| $\Im(\widetilde{\omega})$ $\Delta_0 = 0.8$ | 3.51e-001 | 3.51e-001 | 3.51e-001 | 3.51e-001 | 3.51e-001 | 3.50e-001 | 3.50e-001 | 3.49e-001 | 3.44e-001 |

Another geometrical feature in Fig. 1 for frequencies just below the phase transition happens with zero gap values of 0.8, 0.9 and 1.0 (meV), where the tiny gap is stimulated if $c = 0.2$, meaning that if the zero superconducting gap value is increased closed to the threshold value by 0.1 meV increments. The tiny gap produces a region with only boson supercarriers. It is interesting to notice how the tiny gap in the imaginary part becomes smaller, as $\Delta_0$ decreases. Thus, in Fig. 1 with the inverse strength parameter $c = 0.2$, we observe a tiny gap shaping up in the superconducting phase. This feature in Fig. 1 is supported by a sample of numbers in Table 3.

Table 3: Real & imaginary tiny gap numbers in the RPS for different zero gaps (all values in meV) with dilute disorder and two scattering regions:

| $\omega = \Re(\widetilde{\omega})$ | 7.01e-001 | 8.01e-001 | 8.51e-001 | 9.01e-001 | 9.51e-001 | 1.00e+00 |
|---|---|---|---|---|---|---|
| $\Im(\widetilde{\omega})$, $c = 0.0$ & $\Delta_0 = 1$ | 2.27e-001 | 8.17e-002 | 8.63e-007 | 2.21e-008 | 5.57e-005 | 4.41e-002 |
| $\Im(\widetilde{\omega})$, $c = 0.2$ & $\Delta_0 = 1$ | 2.46e-001 | 1.66e-001 | 9.28e-002 | 5.14e-006 | 6.12e-008 | – – – – |
| $\Im(\widetilde{\omega})$, $c = 0.2$ & $\Delta_0 = 0.9$ | 2.26e-001 | 1.32e-001 | 3.37e-003 | 8.12e-005 | 4.50e-002 | 5.81e-002 |

We remark that the tiny gaps in the cases studied remain positive as can be seen in Table 2, and in the plot extended to negative values in the 0Y axis, to visualize that in the RPS it follows the rule $\Im[\tilde{\omega}] > 0$. Finally in Fig. 1 is noticed that the tiny gap at $c = 0.2$ and $\Gamma^+ = 0.05$ meV, for $\Delta_0 = 1.0$ meV does not display numbers in the normal state (see the last box of line 3, Table 3).

Two additional simulations with a set of parameters $c = 0.2$, $\Gamma^+ = 0.10$ meV, and $c = 0.2$, $\Gamma^+ = 0.15$ meV, i.e., quasi-optimal and optimal disorder, show similar behavior as for the unitary limit [58], with a maximum a zero real frequencies, and an imaginary minimum that decreases as $\Delta_0$ takes smaller values. Figures 2 and 3 show how the zero gap transition smoothly decreases. For example, we notice from Fig. 2 that only for the value $c = 0.2$, $\Gamma^+ = 0.10$ meV, and $\Delta_0 = 1.0$ meV (dark blue plot) there is a sharply minimum but the tiny gap does not shapes up as in Fig. 1. We also observe that the transition point is a smooth function in most cases which says about nonlocality in strontium ruthenate, and changes the position and the slope, as $\Delta_0$ increases. We think, that this feature belongs to triplet superconductors that break the time-reversal symmetry.

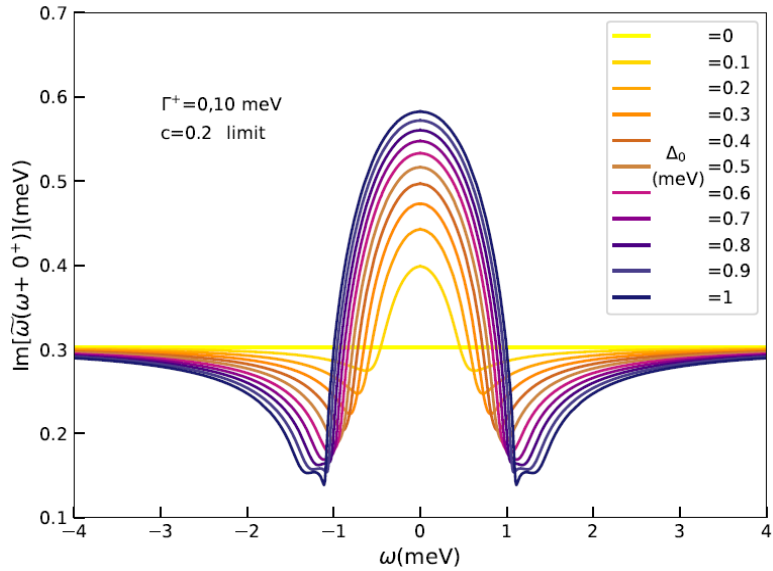

Fig. 2: Evolution of the zero superconducting gap for c = 0.2 and quasi-optimal disorder in strontium ruthenate.

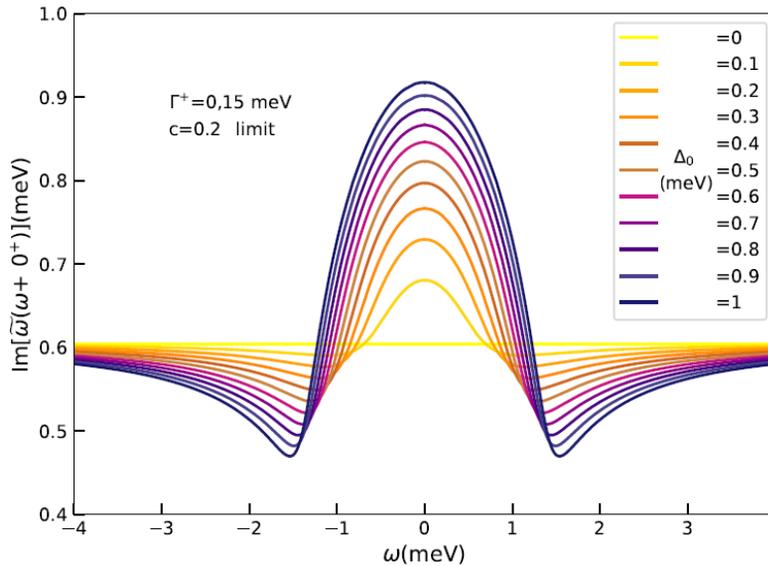

Fig. 3: Evolution of the zero superconducting gap for c = 0.2 and optimal disorder in strontium ruthenate.

### 2. Enriched doped constant superconducting lifetime

In this subsection, Fig. 4 addresses the question if only at the stoichiometric dilute disorder of $\Gamma^+ = 0.05$ meV, a constant superconducting lifetime is settled self-consistently. The question is answered modelling the intermediate regime for strontium ruthenate ($c = 0.4$) and varying the zero gap from the normal state value to the threshold limit, with an enriched stoichiometric disorder given by $\Gamma^+ = 0.20$ meV. Figure 4 shows that in that particular case a robust flat imaginary cross-section develops inside the superconducting phase for the real frequency interval (0.00, 0.36) meV.

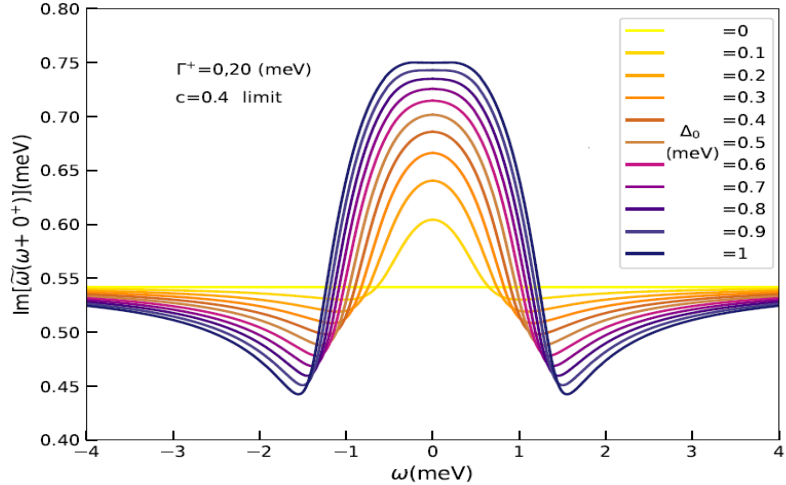

Fig. 4. Evolution of the zero superconducting gap for c = 0.2 and dilute disorder in strontium ruthenate. This plot is flat at real frequencies interval (0.00, 0.36) meV for $\Delta_0$ = 0.9 and 1.0 meV.

Therefore, in Table 4, we present a sample of the numbers calculated. We find that $\Im(\widetilde{\omega}) = 0.749$ meV. This case shows that the real frequency interval with a constant superconducting scattering lifetime decreases as $\Delta_0$ is reduced by 0.1 meV steps. The sample of the values reported in Table 4 supports Fig. 4, and they are calculated using an enriched stoichiometric disorder represented by $\Gamma^+ = 0.20$ meV. To end this section, we think that the last outcome seems to be a general feature of the elastic scattering cross-section in triplet superconductors with a smooth transition where the disorder is induced by stoichiometric Sr atoms.

Table 4: Sample of real & imaginary frequencies, and different zero gap (all values in meV units) for $c = 0.4$ and enriched disorder:

| $\Re(\widetilde{\omega})$ | 1.00e-003 | 2.10e-002 | 4.10e-002 | 7.10e-002 | 1.01e-001 | 1.31e-001 | 1.51e-001 | 1.81e-001 | 3.61e-001 |
|---|---|---|---|---|---|---|---|---|---|
| $\Im(\widetilde{\omega})$ $\Delta_0 = 1.0$ | 7.49e-001 | 7.49e-001 | 7.49e-001 | 7.49e-001 | 7.50e-001 | 7.50e-001 | 7.50e-001 | 7.50e-001 | 7.48e-001 |
| $\Im(\widetilde{\omega})$ $\Delta_0 = 0.9$ | 7.43e-001 | 7.42e-001 | 7.42e-001 | 7.42e-001 | 7.42e-001 | 7.42e-001 | 7.42e-001 | 7.42e-001 | 7.38e-001 |
| $\Im(\widetilde{\omega})$ $\Delta_0 = 0.8$ | 7.35e-001 | 7.34e-001 | 7.34e-001 | 7.34e-001 | 7.34e-001 | 7.34e-001 | 7.34e-001 | 7.33e-001 | 7.28e-001 |

**CONCLUSIONS**

This work aimed at introducing some numerical examples with a constant scattering self-consistent lifetime in the unconventional superconductor strontium ruthenate, and it is inspired in the experimental research work [59]. The singular phase that shows an imaginary elastic scattering cross-section with a single value for an interval of real frequencies in the RPS is found for two scattering limits. The importance of this numerical finding relies on its use, as it justifies the experimental fits of nonequilibrium experimental data.

Thus, in the simulation of Fig. 1 and Table 2, it was analyzed a limit with the strength fixed at c = 0.2, and a stoichiometric dilute disorder $\Gamma^+$ = 0.05 meV (standing in between the unitary and the intermediate regimes). The set of real frequencies in this case is found to be in the interval (0.00, 0.26) meV and the value of the superconducting $\Im(\widetilde{\omega}) = 0.362$ meV. This gives a value of 0.181 meV for the superconducting collisional frequency $\nu$, and a value of 0.764 meV for the inverse scattering lifetime $\tau_s^{-1}$. These three values can be contrasted with those of the normal metallic state (with a Fermi liquid behavior) i.e., the normal phase set $\Im(\widetilde{\omega}) = 0.149$ meV, $\nu = 0.075$ meV, and $\tau_n^{-1} = 0.298$ meV. We notice how the collisional superconducting frequency doubles the normal state collisional frequency for a considerable atomic Sr potential strength $U_0 \gg 1$ despite in both phases the $\tau$ is constant, and the disorder concentration of Sr atoms is diluted.

Second, we also studied a limit with the strength fixed at intermediately regimen, i.e., c = 0.4, and an enriched stoichiometric disorder $\Gamma^+$ = 0.20 meV. A broader (0.00, 0.36) meV interval of real frequencies throws a constant $\Im(\widetilde{\omega}) = 0.749$ meV. The superconducting collisional frequency fixed at $\nu = 0.382$ meV, and a value of 1.528 meV for the inverse scattering lifetime $\tau_s^{-1}$. These three values can be compared in Fig 4 and Table 4 with its normal metallic state where $\Im(\widetilde{\omega}) = 0.524$ meV, $\nu = 0.262$ meV, and $\tau_n^{-1} = 1.048$ meV. We notice how for an enriched stoichiometric $\Gamma^+ = 0.20$ meV, the collisional superconducting frequency increases with respect to the value at dilute disorder $\Gamma^+ = 0.05$ meV of Fig. 1 and Table 2. This finding remarks the role of strontium atoms in the elastic scattering process when disorder is included.

Therefore, we can state that some self-consistent computations using an order parameter with the irreducible representation $E_{2u}$ finds that:

- An state with strength $c = 0.2$ and dilute disorder $\Gamma^+ = 0.05$ meV shows several tiny gaps (Fig. 1 and Table 3), contrasting the unitary Miyake-Narikiyo case with $c = 0.0$ and dilute disorder $\Gamma^+ = 0.05$ meV that shows a single tiny gap.

- The ratio of the normalized ultrasound attenuation in the dilute disorder case is proportional to $\mathbb{C} = 0.390$ (Fig. 1 and Table 2). We find that $\alpha_j(T)\big/\alpha_j(T_c) \sim \tau_s\big/\tau_n = C\big/C' = \mathbb{C} = 0.298\big/0.764 = 0.390 < 1$.

- The ratio of the normalized ultrasound attenuation in the enriched disorder case is proportional to $\mathbb{C} = 0.686$ (Fig. 4 and Table 4). In this case it follows that $\alpha_j(T)\big/\alpha_j(T_c) \sim \tau_s\big/\tau_n = C\big/C' = \mathbb{C} = 1.048\big/1.528 = 0.686 < 1$.

We focused our study outside the c = 0 regime. The set of calculations in the unitary limit varying the zero gap were performed in [52], in order to study other physical properties of strontium ruthenate.

The arguments sketched in the previous paragraphs help to understand the physics of disordered strontium ruthenate, where research about its superconducting nature is still growing despite it was discovered 30 years ago. Some new interesting directions in strontium ruthenate and similar compounds such as strontium ferrite under pressure are mentioned in [61]. In [62-64] and works therein, other properties of strontium ruthenate are discussed. We finally mention a nonlocality discussion of the compounds palladium-cobalt delafossite and strontium ruthenate in [65], and other properties in several families of superconducting unconventional alloys such as heavy fermions and mechanisms for changing Tc with fluctuations addressed in [66,67] respectively.

**Authorship contribution statement**
Pedro Contreras: Conceptualization, Methodology, Software, Investigation, Validation, writing – original draft, Supervision, Writing – review and editing.

**Declaration of competing interest**

The author declare that he has no known competing financial interests or personal relationships that could have appeared to influence the work reported in this paper.